\documentclass{emulateapj}

\usepackage{url}
\usepackage{multirow}
\usepackage{amsmath}
\usepackage{xcolor}

\definecolor{orange}{RGB}{255,127,0}

\tabletypesize{\scriptsize}

\shorttitle{PAPER Power Spectrum over Cosmic Time}
\shortauthors{Jacobs et al.}

\begin{document}

\title{Multi-redshift limits on the 21cm power spectrum from PAPER}
\author{
Daniel C. Jacobs\altaffilmark{1},
Jonathan C. Pober\altaffilmark{12},
Aaron R. Parsons\altaffilmark{2,8},
James E. Aguirre\altaffilmark{3},
Zaki Ali\altaffilmark{2},
Judd Bowman\altaffilmark{1},
Richard F. Bradley\altaffilmark{4,5,6},
Chris L.  Carilli\altaffilmark{7,10},
David R. DeBoer\altaffilmark{8},
Matthew R. Dexter\altaffilmark{8},
Nicole E. Gugliucci\altaffilmark{5},
Pat Klima\altaffilmark{5},
Adrian Liu\altaffilmark{2,11},
Dave H. E. MacMahon\altaffilmark{8},
Jason R. Manley\altaffilmark{9},
David F. Moore\altaffilmark{3},
Irina I. Stefan\altaffilmark{10},
William P. Walbrugh\altaffilmark{9}}

\altaffiltext{1}{School of Earth and Space Exploration, Arizona State U., Tempe, AZ}
\altaffiltext{2}{Astronomy Dept., U. California, Berkeley, CA}
\altaffiltext{3}{Dept. of Physics and Astronomy, U. Pennsylvania, Philadelphia, PA}
\altaffiltext{4}{Dept. of Electrical and Computer Engineering, U. Virginia, Charlottesville, VA}
\altaffiltext{5}{National Radio Astronomy Obs., Charlottesville, VA}
\altaffiltext{6}{Dept. of Astronomy, U. Virginia, Charlottesville, VA}
\altaffiltext{7}{National Radio Astronomy Obs., Socorro, NM}
\altaffiltext{8}{Radio Astronomy Lab., U. California, Berkeley, CA}
\altaffiltext{9}{Square Kilometer Array, South Africa Project, Cape Town, South Africa}
\altaffiltext{10}{Cavendish Astrophysics Group, University of Cambridge, Cambridge, UK}
\altaffiltext{11}{Berkeley Center for Cosmological Physics, UC Berkeley, Berkeley,CA}
\altaffiltext{12}{Dept. of Physics, University of Washington, Seattle, WA}

\begin{abstract}
The epoch of reionization power spectrum is expected to evolve strongly with redshift, and it is this variation with cosmic history that will allow us to begin to place constraints on the physics of reionization.  The primary obstacle to the measurement of the EoR power spectrum is bright foreground emission.  We present an analysis of observations from the Donald C. Backer Precision Array for Probing the Epoch of Reionization (PAPER) telescope which place new limits on the HI power spectrum over the redshift range of $7.5<z<10.5$, extending previously published single redshift results to cover the full range accessible to the instrument.   To suppress foregrounds, we use filtering techniques that take advantage of the large instrumental bandwidth to isolate and suppress foreground leakage into the interesting regions of $k$-space.  Our 500 hour integration is the longest such yet recorded and demonstrates this method to a dynamic range of $10^4$.   Power spectra at different points across the redshift range reveal the variable efficacy of the foreground isolation.  Noise limited measurements of $\Delta^2$ at $k=$0.2hMpc$^{-1}$ and z$=7.55$ reach as low as (48mK)$^2$ ($1\sigma$).  We demonstrate that the size of the error bars in our power spectrum measurement as generated by a bootstrap method is consistent with the fluctuations due to thermal noise.  Relative to this thermal noise, most spectra exhibit an excess of power at a few sigma.  The likely sources of this excess include residual foreground leakage, particularly at the highest redshift, and unflagged RFI.   We conclude by discussing data reduction improvements that promise to remove much of this excess. 
\end{abstract}


\section{Introduction}
The Epoch of Reionization, when the first luminous objects ionized the pervasive cosmological hydrogen, is predicted to be observable in highly redshifted 21 cm radiation.  The Donald C. Backer Precision Array for Probing the Epoch of Reionization (PAPER, \cite{Parsons:2010p6757})\footnote{\url{eor.berkeley.edu}} is a low frequency radio interferometer experiment dedicated to opening this window on the universe.  Challenges include foregrounds which are brighter by several orders of magnitude and long integration times necessitated by the limited collecting areas of first-generation instruments. Direct observation of hydrogen before and during re-ionization is predicted to deliver a wealth of cosmological and astrophysical data, including the nature of the first stellar objects and the timing and rate of galaxy formation. Reviews on the physics of reionization as well as theoretical expectations on the nature of foregrounds may be be found in \citet{Furlanetto:2006p2267,Morales:2010p8093,Pritchard:2012p9555}.  

Other telescopes seeking to measure this signal include the Giant Metre-wave Radio Telescope (GMRT; \cite{Paciga:2013p9943}), the Low Frequency Array (LOFAR\footnote{\url{www.lofar.org}}; \cite{Yatawatta:2013p9699}) and the Murchison Widefield Array (MWA\footnote{\url{mwatelescope.org}}; \cite{Bowman:2013p9950} and \cite{Tingay:2013p9022}).

PAPER is located in the Karoo desert at the site of the South African portion of the future Square Kilometer Array\footnote{\url{skatelescope.org}} and has doubled in size on a yearly basis since 2009; science-grade observations have been made with each stage of the build-out.  

Here we report on deep integrations made with a 32 element array in 2011, first described in \cite{Parsons:2014p10499}, hereafter P14.  Our data reduction pipeline was described in detail in P14, where the same methods were used to give the deepest limits yet on the HI power spectrum in the presence of bright foregrounds at redshift 7.68.  However, the Epoch of Reionization signal is expected to evolve strongly with redshift. In fact, it is this signature variation which will distinguish it from foregrounds and it is this variation with cosmic history that will allow us to begin to place constraints on the physics of reionization \citep{Pritchard:2008p8123,Pober:2014p10390}.  Therefore, while a detection of the 21 cm signal at even a single frequency would be a tremendous breakthrough, analysis techniques must be developed to capitalize on the wide bandwidths of the current generation of high-redshift 21cm telescopes.  Using the same data set as P14, this paper presents improved upper limits on the HI power spectrum over the redshift range $10.7>z>7.2$.  In Section  \ref{sec:observations} we summarize the observations, in Section \ref{sec:obs_meth} review the reduction methodology, we present the new upper limits in Section \ref{sec:results}, and in Section \ref{sec:conclusion} we offer conclusions and discussion of future work.

\section{Observations}
\label{sec:observations}
The work here follows the same basic procedure and uses the same underlying data set as P14. Here we provide a quick summary and refer the reader to P14 for a more in-depth discussion.  A general overview of the PAPER system can be found in \cite{Parsons:2010p6757}, calibration of the primary beam in \cite{Pober:2012p8800}, and imaging results in \cite{Jacobs:2011p8438,jacobs:2013b} and \citet{Stefan:2013p9926}.  Sensitivity analysis described in \cite{Parsons:2012p9028} revealed that for the low gain elements employed by PAPER, a highly redundant ``grid'' type arrangement offers a significant sensitivity boost extending the performance of the compact designs motivated by \citet{Morales:2004p2494,Bowman:2006p1887,Lidz:2008p8251}.  In most interferometers the locations of the antennas are optimized such that each baseline samples a different Fourier mode of the sky; this is the ideal case for reconstructing images where each mode contains different information.  For a power spectrum measurement the key metric is sensitivity per mode, rather than number of modes.  A grid configuration allows many samples of each cosmological mode, to be averaged to a high sensitivity before being combined with other Fourier modes.  The PAPER South Africa 32 antenna deployment (PSA32) was arranged in a 4$\times$8 grid with a column spacing of 30m and a row spacing of 4m.  In our analysis, as in P14, we include only the three shortest types of spacings where the reionization power is expected to be brightest. This selection includes those between adjacent columns and within at least one row of each other, a selection containing 70 $\sim$30m-long baselines.  We will use these baselines to make a one dimensional estimate --spectral line of sight only-- of the HI power spectrum.

Observations spanning the band between 100 to 200-MHz ($13.1>z>6.1$) were recorded at a resolution of 48kHz and 10.7s  beginning Dec 7, 2011 and ending March 19, 2012 (with some down-time for maintenance) giving a total of 513 hours over 92 nights.  Within this set we included observations in the LST range 1h - 9h where the sky dominated system temperature is at a minimum.  Note that this LST range is slightly shorter than in P14, which extended to LST of 12 hours. These last three hours were found to contribute minimally to increasing sensitivity while being dominated by bright galactic foreground emission and so have been excluded here.

\section{Reduction}
\label{sec:obs_meth}
Here we summarize our data reduction steps; for more details see Section 3 of P14.  In summary, we use 70 nearly identical baselines to make a 1D estimate along the spectral or line-of-sight direction of the reionization era HI power spectrum.  All processing, save calibration, and  the final cross-multiplication step treats each baseline as independent. Foregrounds and interference are removed on a per-baseline basis with no a-priori sky model using signal processing techniques and a physical model of the array. In the final cross-multiplication step, the last layer of systematics is estimated and removed by projecting non-physical correlations between baselines.
\subsection{Delay/Fringe-Rate Filtering: Averaging and Foreground Removal}
\label{sec:transforms}
In several stages throughout the analysis process we take a 2D Fourier transform of the visibility spectra $V(\nu,t)$ into ``delay/fringe rate'' space where delay is the Fourier dual to frequency and similarly fringe-rate for time.  In this space, smooth spectrum sources are physically localized to delays shorter than the light travel time length of the baseline and fringe rates shorter than the sidereal rotation rate of the tip of the east-west component of the baseline vector. Sources at the horizon, in the direction of the baseline vector, have the longest delays, while fringe rates are highest where the celestial equator crosses the horizon.

 In this Fourier space, sources are highly localized with deviations from a flat spectrum manifesting as a slight dispersion. The spectrum sampling function, which is uneven due to flagging of interference takes the form of a convolution by a point-spread-function (PSF) in the same way an imperfect sampling of the $uv$ plane gives rise to the angular PSF of an interferometer.  If enough data is missing this PSF can cause smooth-spectrum sources to leak beyond the ``horizon'', the light travel time limit.  To account for this, we use a CLEAN like, iterative, peak-finder and subtraction algorithm which is limited to finding peaks within the physically allowable ranges of delay and/or fringe-rate. In this case, the 1D ``CLEAN'' beam is the Fourier transform of the spectral or time sampling function \citep{Parsons:2009p7859}.

The data analysis pipeline essentially consists of iterative application of the delay or fringe-rate transform process, with an ever tightening allowable number of modes, interleaved with stages of averaging (time, frequency, night), before finally computing a power spectrum.  This final step takes advantage of the redundant baselines to make an unbiased power spectrum estimate by cross-multiplying identical baselines and then averaging the power spectrum modes. By not combining baselines until the last step and by assuming that the line of sight transform is well approximated by the delay transform we dramatically simplify the analysis procedure to a series of signal processing steps. It also obviates the need for precision sky and instrumental beam models which are required by imaging arrays.  

This isolation of foregrounds to a region below a line-of-sight $k_\parallel$ mode that increases with baseline length is also the much discussed ``wedge'' \citep{PhysRevD.90.023018,PhysRevD.90.023019,Thyagarajan:2013p10039,Pober:2013p9942,Trott:2012p10466,Morales:2012p8790,Parsons:2012p8896,Vedantham:2012p10297,Datta:2010p8781,Parsons:2009p7859}. As we only have three baseline types of nearly the same length the wedge manifests as a single value of $k\sim k_\parallel$, below which foregrounds are expected to dominate.  The details of each application of the delay/fringe-rate transform will be laid out in the following sections as we provide a brief walk-through of the processing pipeline.

\subsection{Selection of Redshift Bins}
The redshift bins for which we have computed power spectra --shown in Figure \ref{fig:bins}-- have been selected from the available bandwidth using two criteria: minimizing covariance between redshifts and avoiding missing spectral data. We minimize covariance between adjacent redshift bins by limiting overlap of each bin to the spectrum at the outer half of the bin channel range which, as will be described in section \ref{sec:power_spectrum}, is significantly down-weighted by the use of a Blackman-Harris window.   Avoiding missing data is important because the power spectrum method, developed in P14, which leverages the redundant baselines to estimate non-sky covariance and project out these contaminated modes is particularly sensitive to missing data -the inverse of the covariance is not defined- so we also select only redshift ranges with channels that have no missing data over the entire sidereal period (see the dotted line in Figure \ref{fig:bins}).  With these constraints we arrive at the redshifts 10.3, 8.5,7.9 and 7.45.  For the purposes of comparison with P14 we also include the redshift 7.68 bin.  \citet{Dillon:2013p10497} describes a pseudo-inverse method for handling the covariance introduced by missing data. Though not implemented here, this method is in the process of being adapted for use in highly redundant analysis \citep{PhysRevD.90.023018,PhysRevD.90.023019}.  

The spectral width of the redshift bins is dictated by two competing needs. First, sensitivity and foreground reduction both benefit from wider bandwidths. On the other hand, power spectrum measurements which trace the evolution with redshift require smaller bandwidths. Often, the evolution scale is taken to be on order $\Delta z \approx 0.5$, which translates to a bandwidth limit $B<0.5 (f^2/1421)$ MHz. To balance these two competing constraints we choose a single spectral bandwidth of 20MHz, weighted by a Blackman-Harris window for an effective width of 10MHz, or $\Delta z=0.5$ at $z=7.45$ ranging up to a $\Delta z = 0.86$ at redshift 10.3.

\subsection{Initial Averaging}
  First, the raw data are down-selected to just the 70 $\sim$30m long baselines described in Section \ref{sec:observations}.  
   The visibilities are then compressed in the frequency and time directions by removing delay modes and highest frequency fringe rates corresponding to a 300m baseline (the longest baseline in the array).  This filtering\footnote{See Appendix A of P14} is done in tandem with a radio frequency interference (RFI) flagging step, using the residuals which have had bright sky-like signals removed to flag 4$\sigma$ deviations before feeding the flags back into another iteration of the compression step. This mitigates the effects of bright, narrow band, interference  being scattered into higher delay modes (where only the reionization signal is expected to be found) and results in a time and frequency bin of length 40s and width 492.61kHz. 
  This process reduces the data volume by a factor of $\sim$20, or roughly an order of magnitude improvement on traditional time and frequency averaging which in this array would be limited to ~100kHz and 10s to avoid averaging away fringes.
  
\subsection{Calibration}
We model the gain as a per-antenna amplitude and a phase slope -physically a single time delay- and single real, low order polynomial passband for all antennas.  Because the array samples correlations redundantly, the relative calibration between antenna is numerically overdetermined and tractable as a linear algebra problem \citep{Liu:2010p10391}.  As described in P14, we compute the ratio between redundant baselines, fit for a gain and phase slope and then algebraically solve for a per-antenna solution.  Using this method we have avoided calibrating each channel independently to preserve as much frequency variation as possible.   These solutions vary little over the three month observing period, exhibiting less than 1\% r.m.s. variation. A single solution derived for the Dec 7 data set is used for the entire observing run. Time and frequency variation of redundant solutions is explored in general in \citet{Zheng:2014p10467}.  

Relative calibration allows the gains of different antennas to be calibrated against one another.  However, to obtain a correctly normalized power spectrum, it is necessary to set a flux scale for which we need a flux measurement of a calibration standard.  For this we form a beamform on a known, bright calibration source.  By itself, the redundant calibration does not contain enough phase information to phase coherently to a sky location, there remain two free phase parameters which cannot be solved by redundancy alone.   We find these by fitting a model of Pictor A, Fornax A, and the Crab Nebula during a time when the sky is dominated by these three sources while marginalizing over the unknown\footnote{Unknown in the sense of a joint uncertainty in source flux and primary beam pattern. Though the fluxes of these sources is in some cases fairly well constrained, fluxes at 150MHz are still fairly uncertain as is the PAPER beam in those directions.} apparent flux ratio between the three sources. With the delays in place we are now able form a beam on Pictor A and (for each channel) set the overall amplitude to the calibration value of 382 (f/150MHz)$^{-0.77}$Jy found in \cite{jacobs:2013b}.

 \begin{figure*}
\centering
\includegraphics[width=\textwidth]{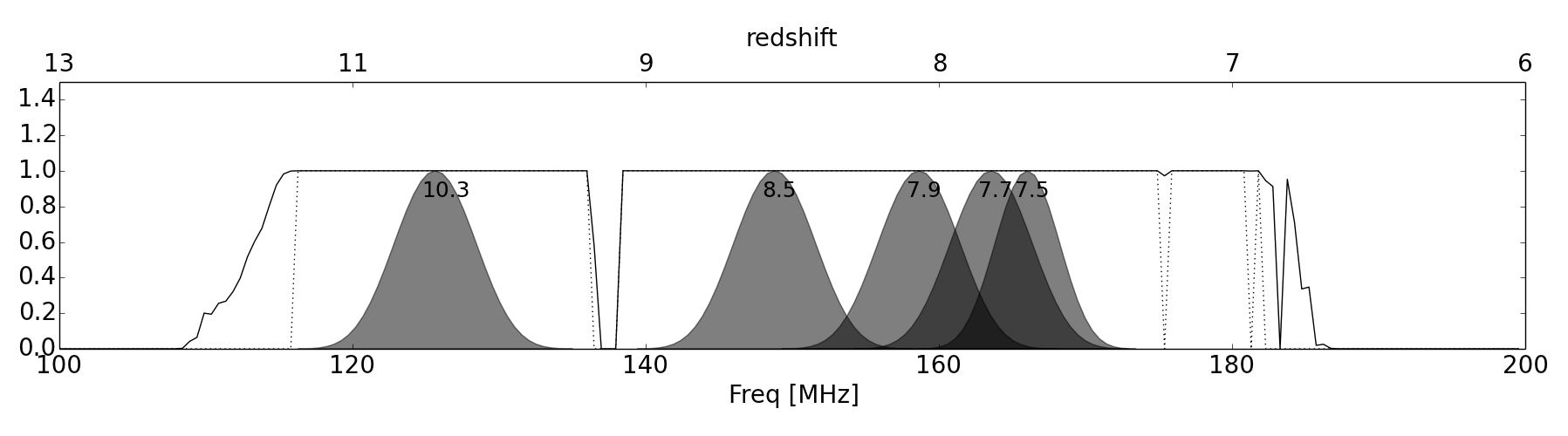}
\caption{\label{fig:bins} The average amount of data remaining after interference flagging over the 3 month period between Dec 2011 and March 2012 (black line) is quite high.  Redshift bins (in grey, redshift center indicated with text label) are chosen to include spectral channels with uniform weight, i.e. no missing channels while maximizing coverage over the band.  Redshift 7.68 is included for comparison with P14.  Channels with no missing data are indicated by the dotted line, visible at the edge of flagged channels. Each redshift bin is 20 MHz wide, but weighted by a Blackman-Harris window function which heavily down-weights the outer 10MHz for a Noise Equivalent Bandwidth of 10MHz.  The interference is almost exclusively dominated by two features: ORBCOMM satellites at 137MHz and an unidentified intermittent line emitter at 175MHz. The roll off at 115MHz is due the rising noise at low frequencies being incorrectly flagged as interference.  }
\end{figure*} 

\begin{figure}
\centering
\includegraphics[width=\columnwidth]{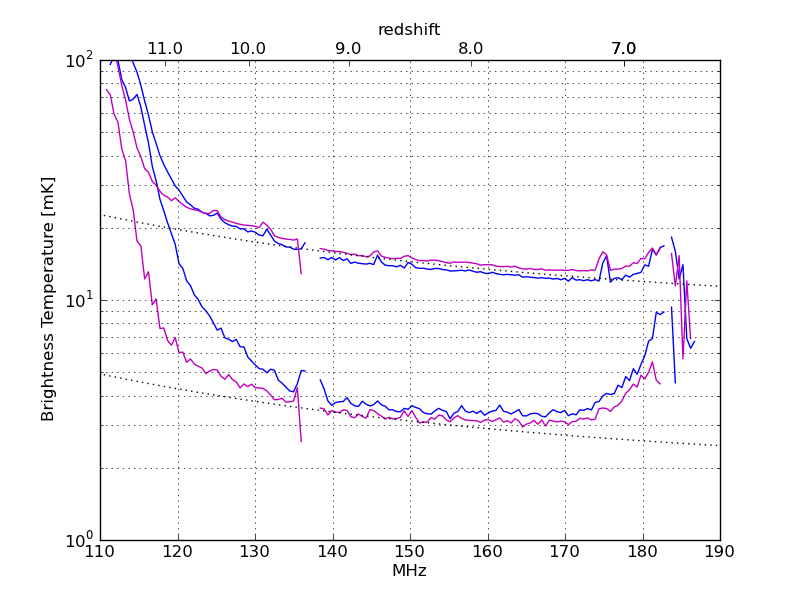}
\caption{\label{fig:noise} Root mean square (rms) noise before and after integrating to 13 minute LST bins (top and bottom sets of curves) indicating the rising significance of foregrounds as we take the last deep integration step. Noise is estimated by differencing adjacent frequencies (magenta) and between redundant baselines (blue), compared with an estimate of the theoretical noise level with a $T_{sys}$ of 550K (dotted). The top three lines show noise after filtering foregrounds and binning into 40s long sidereal bins.  At this noise level the frequency and baseline differences are roughly similar, both demonstrate the same small bumps of increased noise due to interference flagging and are consistent with noise over much of the band.  The bottom three lines show the noise level after integrating up to the maximum fringe rate of 789s. The step between the top and bottom sets is the last coherent integration with noise largely decreasing by the expected factor $\sqrt{789/40}$ except in the difference between baselines (blue) which demonstrates a clear excess at all frequencies, particularly above redshift 10. With this last large jump in sensitivity we are now seeing the slight dominance of baseline covariance over the spectral derivative rms. }
\end{figure}

\subsection{Foreground Filtering and Night Averaging}
Foregrounds are filtered from the calibrated data by removing all bright delay components with light travel times less then the baseline length. Where during the previous compression step a liberal horizon of 300m (1800ns, much longer than the 30 meter baselines under study)  was used to calculate the window size, we now choose a window corresponding to the 30m baselines under study with an extra buffer of 15ns to provide a slight buffer against the 1/B$\sim$12ns resolution of the delay spectrum.  The broadband delay spectrum model is iteratively built then subtracted from the data leaving residuals which we carry into the next stage. Next, a four hour long running mean is subtracted. This removes excess correlation due to cross-talk in the analog signal chain. The residuals are then flagged once more for RFI before the 92 nights of data are averaged into 40 second long local sidereal time (LST) bins, which as PAPER is a drift-scanning instrument, are equivalent to bins in Right Ascension (Declination is fixed at -30\arcdeg).  During averaging we found that some LST bins were dominated by a small number of exceedingly bright samples lying well outside the rest of the gaussian distributed data. To compensate we filter the 10\% brightest samples in each bin.  The source of these outliers is not known,  a likely possibility is an instability in the analog signal chain stimulated by weather or bright interference, a circumstance that has since been observed in later seasons.

Though the frequency and repeated nightly observations have been averaged to their maximum extent, at 40s integrations the time axis has yet to be fully exploited.  Sky-like fringes on a 30m baseline rotate much slower than 40 seconds.  Performing a final fringe-rate filter, limiting to fringe-rates expected on a 30m baseline (down from 300m in the last iteration), we arrive at a data-set averaged to 789s, the maximum possible while still maintaining fringe coherence.    The root mean square of the residual signal (seen in Figure \ref{fig:noise}) at the end of this process is close to the 3mK level expected given the total integration time and  system temperature.

\subsection{Power Spectrum}
\label{sec:power_spectrum}
The output of the above steps is eight sidereal hours of calibrated, foreground filtered visibility data averaged over 92 nights. The power spectrum is estimated in the delay spectrum of  a 10MHz bandwidth range centered on the redshift of interest. To preserve the isolation of any foregrounds which remain, we increase the spectral range by 5MHz on each side and multiply by a Blackman-Harris window thus providing a much higher dynamic range delay spectrum point spread function. 

This leaves us with 40 delay samples on each of 70 baselines which are divided into three redundant groups. Within these groups we cross correlate delay spectra between different redundant baselines.  The cross multiplication of the same delay modes between different redundant baselines provides an unbiased estimate of the power spectrum.  These ``sky-like'' correlations should be identical between all redundant baselines to within the level of the noise, while all other cross multiplications between delay modes should not be correlated between different baselines. In practice  these non-sky-like modes do occasionally have significant power which leaks into the correlations which sample the power spectrum.  These are removed by iteratively dividing the covariance into a model of systematics and a model of sky-like emission and then projecting out large residual modes. This is done by dividing the baselines into different groups such that all cross-multiplications are done without introducing noise bias.  For more see Appendix C of P14.

The residual elements of the correlation matrix corresponding to cross-multiplication of matching delay bins between different baselines are all un-biased samples of the power spectrum. To estimate the final power spectrum and its uncertainty we compute the mean and variance of many random randomly-selected subsamples, sampling the dimensions: sidereal time, redundant baseline pair, and delay sign\footnote{As visibilities are complex, both the positive and negative delays  carry  information. Physically the two signs correspond to the two halves of the sky.}. For simplicity, measurements are averaged with equal weights.

\begin{figure*}
\centering
\includegraphics[width=0.9\textwidth]{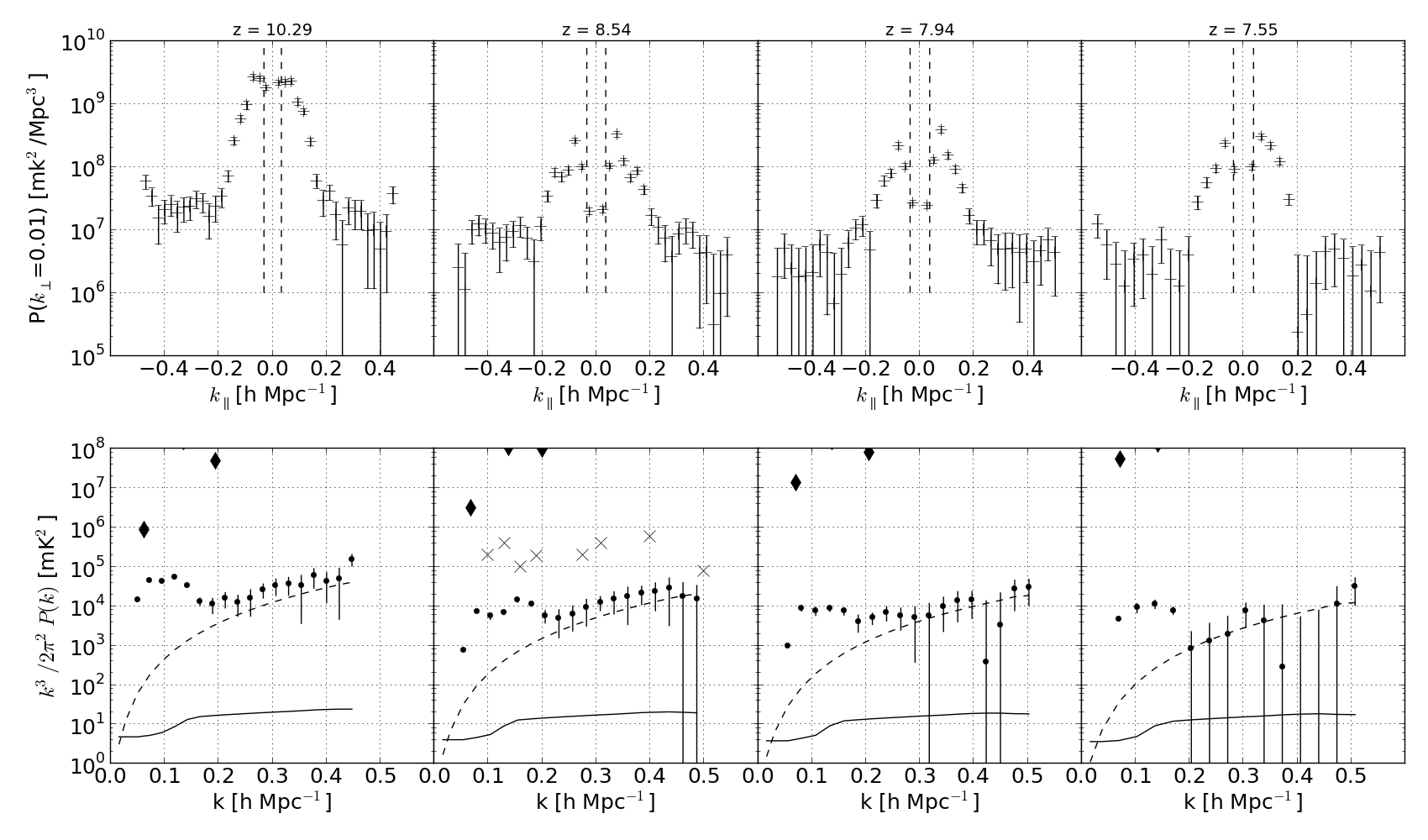}
\caption{ \label{fig:pspecs} PAPER power spectra at four redshifts. On the top, $P_k$ spectra (a simple units change from the raw delay spectrum) provide a useful diagnostic on foreground rejection, while on the bottom we plot in $\Delta^2$ cosmological units with 1 $\sigma$ error bars.    Redshift increases left to right. All spectra have an effective bandwidth of $\pm$10MHz  covering the redshift span $\Delta z\approx$1421MHz ($B/f^2$) which ranges from 0.8 at $z=$10.3 to 0.45 at $z=$7.4.  The noise curve (dashes) is calculated using the method described in \cite{Pober:2013p9581}  and indicates the 1$\sigma$ confidence bounds on data points consisting purely of noise; 65\% of uncorrelated noise like data points will lie below the curve. However, note the caveat that due to the weighting of the delay transform, adjacent $k$ bins are 50\% correlated.  See \cite{Pober:2014p10390,Pober:2013p9581} for a discussion of the approximations made in those calculations. The black line is a fiducial model at 50\% ionization \citep{Lidz:2008p8251}.  GMRT points from \cite{Paciga:2013p9943} indicated with 'x's and MWA points (also using 32 antenna) are black diamonds \cite{Dillon:2014p9788}.   }
\end{figure*}

\begin{figure}
\centering
\includegraphics[width=0.9\columnwidth]{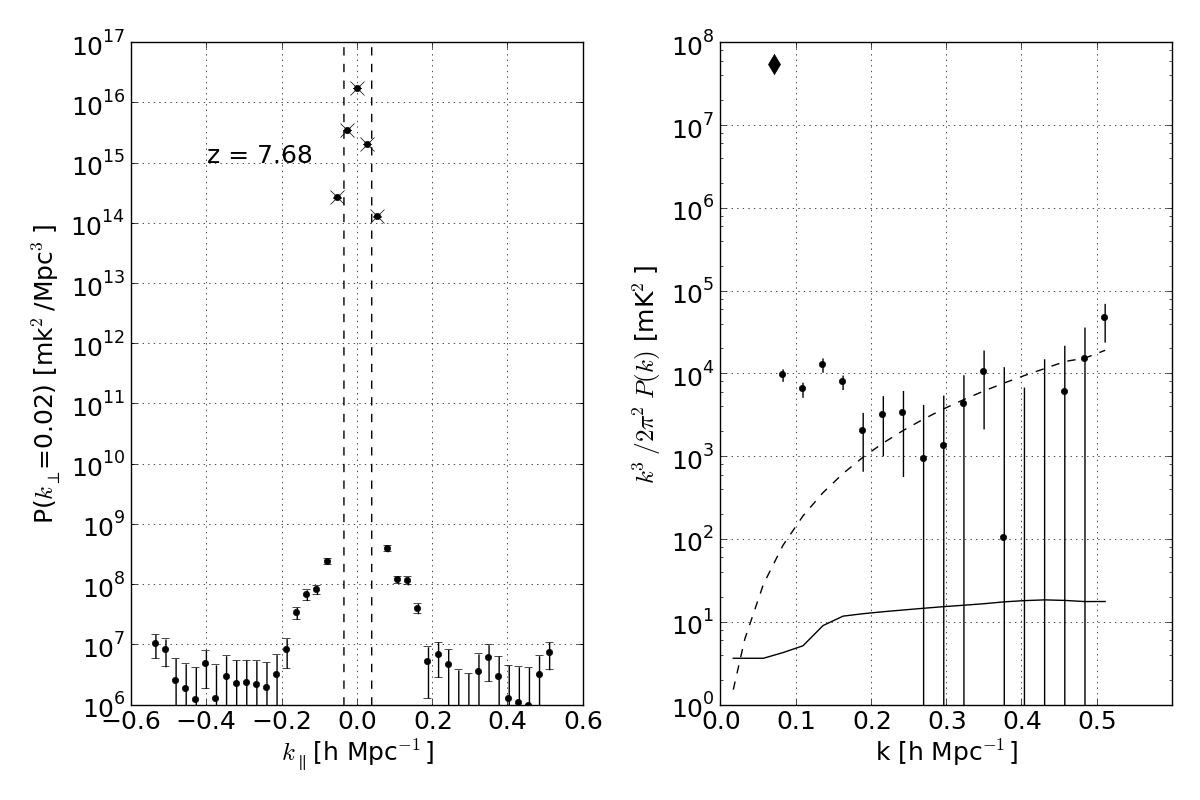}
\caption{\label{fig:P14compare} The redshift 7.68 bin has been reprocessed for comparison with the P14 result.  See Figure \ref{fig:pspecs} for a detailed description of curves. Compare with Figure 6 of P14. Note that the LST range processed here is 3 hours less than P14 for a small reduction in sensitivity but large reduction in foreground. The data shown here was processed exactly as the data in Figure \ref{fig:pspecs}, the foregrounds have been filtered. However the power spectrum shown in P14 added back in the narrow band delay spectrum points (the central five bins inside the horizon which is indicated by dashed lines) for comparison purposes we have added back in these points from that paper. They are marked with '+'s.  }
\end{figure}

\section{Results}
\label{sec:results}
\subsection{Foreground Filtering and Noise Levels}
\label{sec:noise}
The root-mean-square of differenced visibilities measures how well we've removed foregrounds.  In Figure \ref{fig:noise} we examine $T_{rms}$ as calculated by differencing between adjacent channels and between redundant baselines.  The square of this difference is then averaged over the full LST range probed by the power spectra (1h-9h). Note that this operation combines all LST modes as though they were repeated measurements of the same sky when in practice the power spectrum values are averaged in quadrature over different LST bins. This provides a lower bound on the sensitivity possible in the data set and a sensitive probe on foreground residuals or other sources of excess power.  Using all the data in this way, we expect to illuminate rare outliers that effect the bottom line.

\subsubsection{The Theoretical RMS Noise Level}

The expected noise level over this time range depends on the number of samples in each lst bin and the system temperature in that bin. System noise comes in part from the bright radio sky and part from amplifier noise in the receiver. As well as the channel width $B$ and the integration time $t$

\begin{equation}
T_{rms}  =\frac{1}{N_{\textrm{lsts}}}\sum_{\textrm{lst}} \frac{T_{sky}(lst) + T_{rcvr}}{\sqrt{2BtN(lst)}} 
\label{eq:Trms}
\end{equation}

Where $T_{sky}(lst)$ is modeled by the convolution of the global sky model \citep{deOliveiraCosta:2008p2242} with the PAPER beam \citep{Pober:2012p8800} which for the LST range here averages to 250 $(\nu/160)^(-2.6)$~K.

One of the strongest influences on the sensitivity of wide-field, low frequency instruments like PAPER or the MWA is the number of samples local sidereal time bin available in an observing year.   After removing data containing Sun and the Galactic center, both of which are sources of strong foreground emission, the amount of time required to record a deep $\sim$1000 hour observation can stretch into multiple years.  Some LST bins in this range will never be sampled more than a few times per year.  This observation, which ran from December to March, measured samples over half the night sky. However only the range 6h42 to 9h was measured every night to reach ``full'' sensitivity of the three-month-long observing season.  Integrating equation \ref{eq:Trms} over this sampling function we find a $T_{rms}$ noise level equivalent to 40 nights of integration. 

Using the variance of a 20 day subset, P14 found the system temperature for this LST range to be 550~K at 160~MHz, while the sky model suggests a $T_{sky}$ of 250~K.  This level of sky temperature is confirmed by the fit of the theoretical noise level in Figure \ref{fig:noise}.   The sky portion of the noise has a significant slope. Ascribing more then 250~K of the total noise to the sky results in a significantly steeper theoretical $T_{rms}$ in Figure \ref{fig:noise}. Taken together this evidence points to a receiver noise temperature of 300~K. 

In summary we find that the noise level of this data set is well modeled by a 550~K system temperature which is nearly equal parts sky and receiver and when integrated over our lst sampling function is equivalent to 40 days of data.

\subsubsection{Analysis of the RMS Noise Level}

Here we have estimated the residual noise after foreground filtering and LST averaging, both before and after the last ``coherent'' integration step, when fringe rate filtering increases the effective integration time from 40 to 789 seconds.  
The final fringe-rate step is the last coherent average of the processing routine and represents the last large gain in sensitivity; all following averaging steps are done incoherently on the square of the visibilities where sensitivity is gained at a much slower rate.  The noise level of the 40s average (top curves of Figure \ref{fig:noise}) is consistent between the two methods of noise estimation and with the predicted level. 

After the last stage of coherent integration (from the top set of curves to the bottom set), where the integration increases from 40 to 789 second integrations.  At this deepest level of integration the noise levels rise slightly above the expected decrease of $\sqrt{789/40}$. In both cases the noise curves deviate significantly from the theory at the edges of the band. This is mainly due to our neglect of variable sensitivity across the band (e.g. see Figure \ref{fig:bins}) in calculating the theoretical curve; note that the band edges still integrate down between the two steps. However it does not integrate down perfectly, residual levels are apparent across the entire band with larger excesses at the band edges.  The largest excess is noted in the rms difference between redundant baselines.  Lack of agreement between supposedly redundant baselines is suggestive of a breakdown in our assumption of spectrally and temporally smooth calibration. It is also suggestive of the excess covariance between redundant baselines which we remove in section \ref{sec:power_spectrum}, a hypothesis also supported by the high levels of residual foregrounds in the redshift 10.5 bin where the baseline difference r.m.s. is highest.

\subsection{Power Spectra}
\label{sec:pspecs}

In  Figure \ref{fig:pspecs} we show the  power spectra at different redshift bins --showing both the spherically averaged power spectrum $P(k)$ and the volume weighted  $\Delta^2(k)\equiv\frac{k^3}{2\pi^2}P(k)$.  The $P(k)$ spectrum averages the three baseline types but preserves the positive and negative delays; it is essentially just the average delay spectrum scaled to temperature and cosmological units. Note that due to the choice of window function, the adjacent $k$ bins correlate at the  50\% level. In Figure \ref{fig:slices} we see different $k$ mode slices as a function of redshift,  all plotted with $1\sigma$ error bars derived from the bootstrap process and compared with the theoretical noise level. Table \ref{tab:data} we list the data plotted in these slices.   

The theoretical noise power spectrum (dashed line in Figures \ref{fig:pspecs} and \ref{fig:P14compare} and grey region in Figure \ref{fig:slices}) is estimated using the method described in \cite{Pober:2013p9581}, assuming a system temperature of 550K and the observing scheme described in Section \ref{sec:observations}. In general the theoretical error bars are smaller than those derived by the bootstrap.  This is most likely due to the use of uniform weighting when combining power spectrum measurements, rather then the inverse variance weights assumed in the sensitivity calculation.

 In all figures the noise levels are plotted for comparison with the power spectrum values rather than the error bars, i.e. 68\% of uncorrelated, noise-dominated points should lie below the line or within the grey region.    Points where the error bar does not cross zero are inconsistent with noise at $>68$\% these are listed as ``detections'' in Table \ref{tab:data}.   

If our measurements were completely noise dominated, $68\%$ of the data points ought to include zero within their error bars.  In Table \ref{tab:data}, we divide the points into ``Detections" (power spectrum values smaller than $1\sigma$ errors), and ``Upper limits" (measured values larger).  With only one exception, this scheme yields the same results whether we use the theoretical errors or the bootstrap errors.  Under this scheme, we would expect a noise dominated signal to yield around eight ``upper limits'' with an equal number of positive and negative data points. Here we find only two upper limits while the rest we classify as low significance detections of excess power. In general, most detections are within 1 - 2 $\sigma$ but are all positively biased,  with the significance rising steeply at high redshifts and low $k$s.

\subsection{The Excess Power}

As these measurements have a maximum possible sensitivity level which is still two orders of magnitude away from theoretical reionization levels we consider the likelihood that these detections are evidence of reionization to be small. Far more likely is that we are limited by a residual foreground or systematic arising from approximations in the data analysis.  The statistical significance of the residual --2$\sigma$ after averaging over the entire season-- is low enough that it is not possible to break it down along likely axes like sidereal time or time of day to probe its origin.  Such studies must await higher sensitivity measurements with more antennas.  However, there are several additional features that seem to suggest an excess not caused by high redshift HI.

The spectral signature of the measured excess is far more similar to a foreground residual then to theoretical HI models.
The predicted signature of the HI signal during reionization is one of a relatively short burst of power at small $k$ values as the ionized bubbles are briefly at their largest, before the signal from the ever increasing ionized medium drops sharply \citep{Pritchard:2008p8123}. In contrast, this observed residual is relatively constant with redshift while rising at both ends of the band, mimicking the RMS noise curve in Figure \ref{fig:noise}.  This suggests that the accuracy of the foreground filter decreases towards the edges of the band.  Recall that the wide-band foreground model we have subtracted was built by weighting the entire band by a Blackman-Harris window function which provides a much higher isolation of foregrounds inside the horizon. The tradeoff that is made is that data in the outer parts of the spectrum are heavily down weighted.  For example data at 126MHz are down-weighted $\sim$75\% compared to 150Mhz.  This amounts to making an assumption about the spectral smoothness of the data, namely that 126MHz can be mostly modeled by extrapolation from 150MHz.  The precision of this subtraction can be judged by examining residuals inside the horizon in the $P(k)$ power spectra plotted in the upper part of Figure \ref{fig:pspecs}\footnote{Note that this notch is not present in Figure \ref{fig:P14compare} because the foregrounds have been left in for comparison with P14.}.  The relative depth of the redshift 10.5 P(k) foreground residual is substantially shallower then at 7.55 and is generally correlated with the amount of residual found at higher $k$s.

\begin{figure*}
\centering
\includegraphics[width=0.48\textwidth]{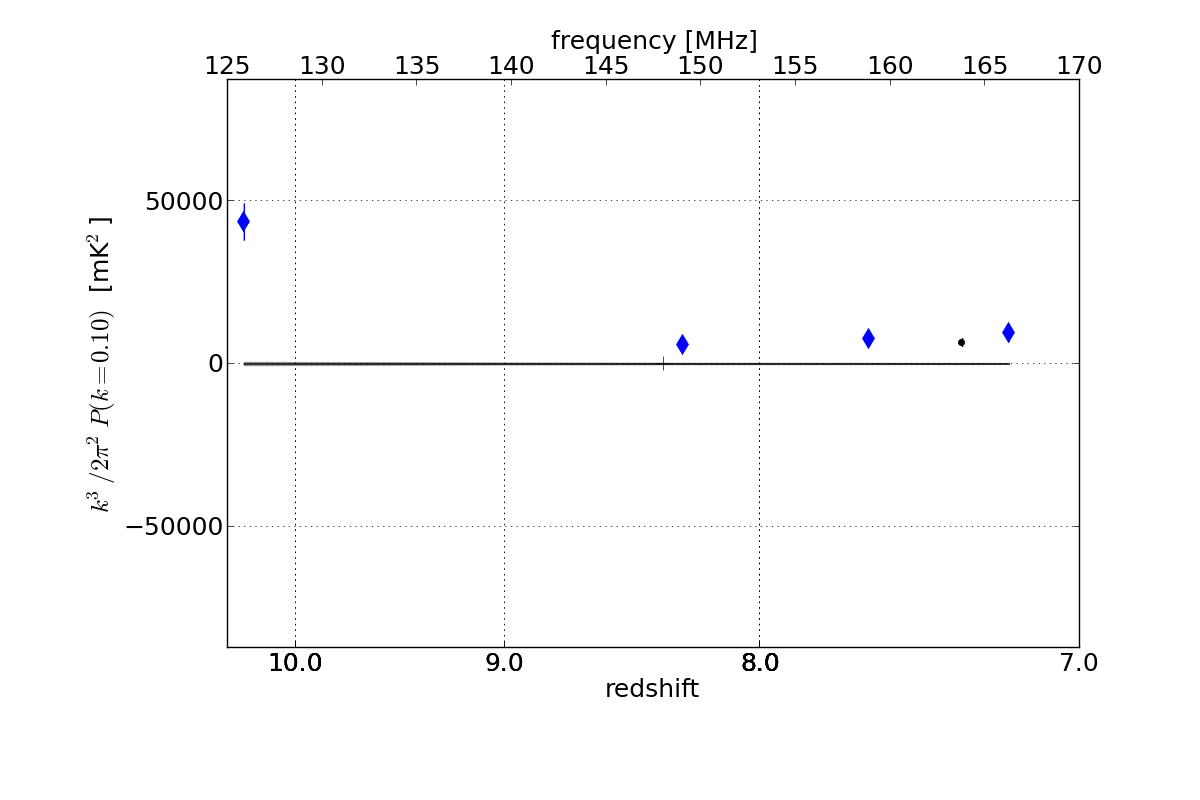}
\includegraphics[width=0.48\textwidth]{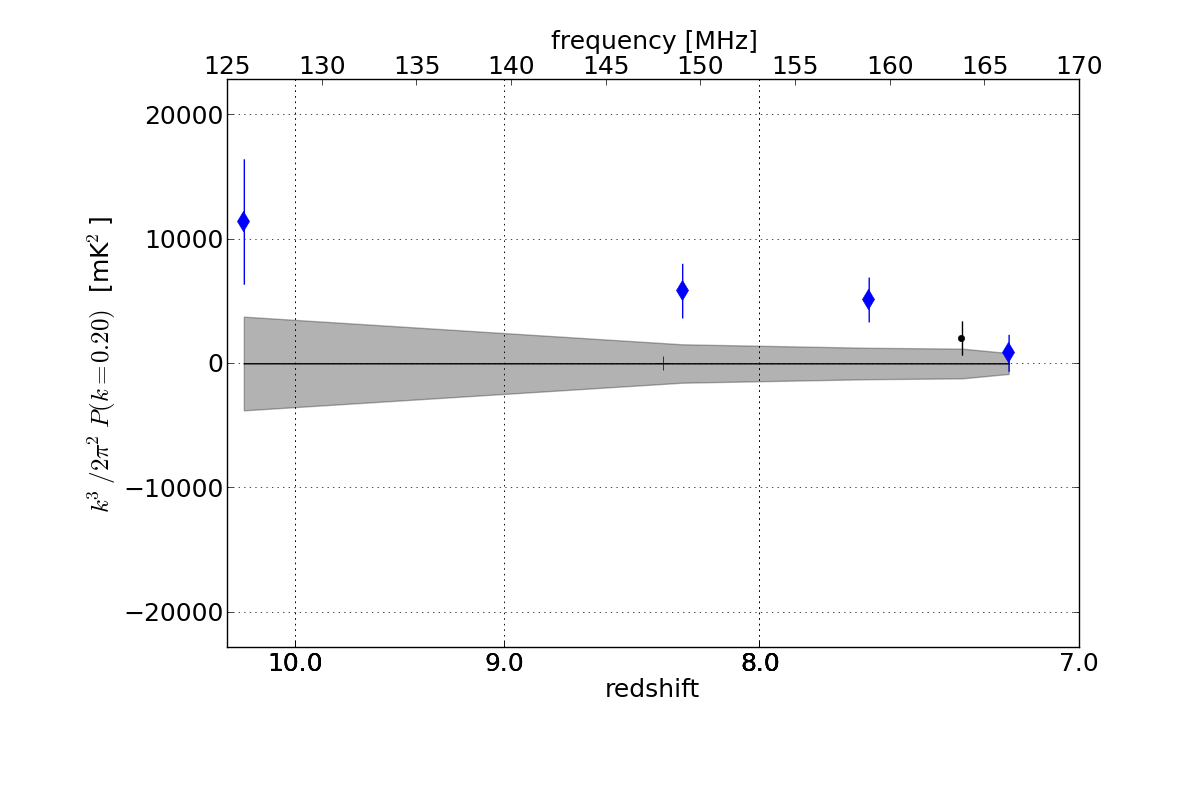}
\includegraphics[width=0.48\textwidth]{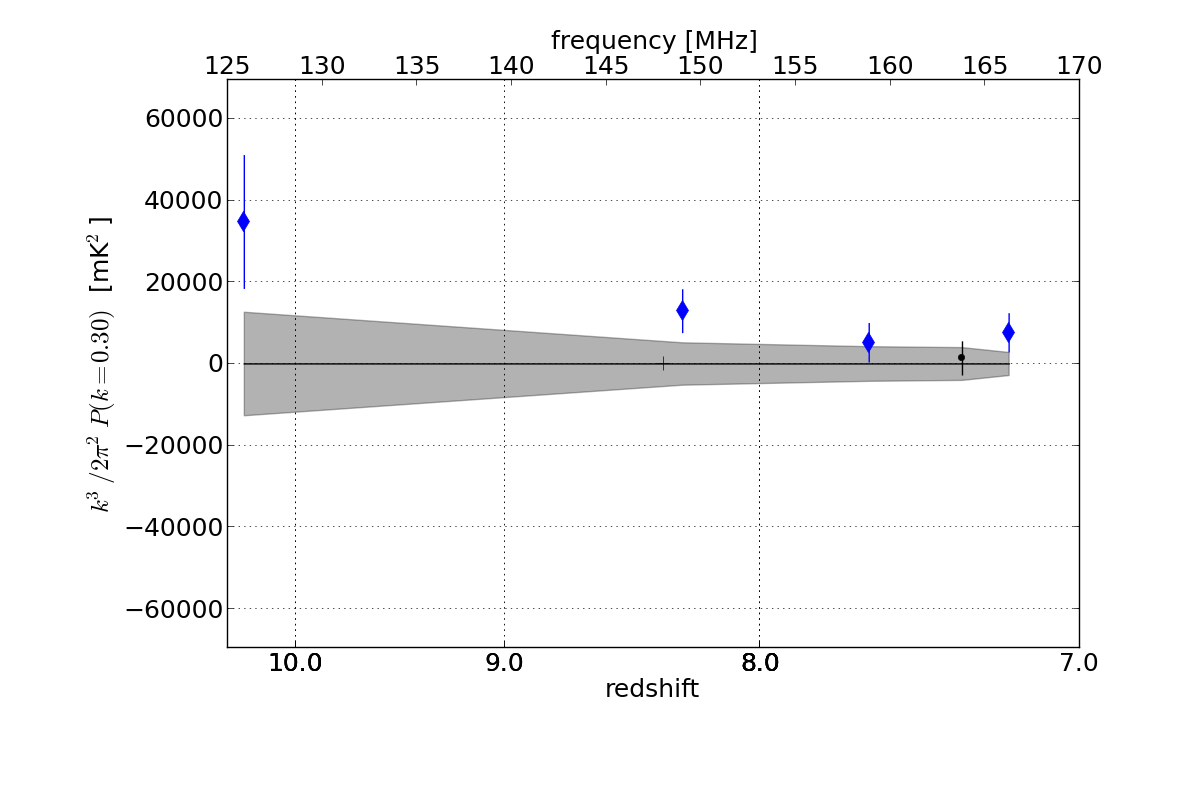}
\includegraphics[width=0.48\textwidth]{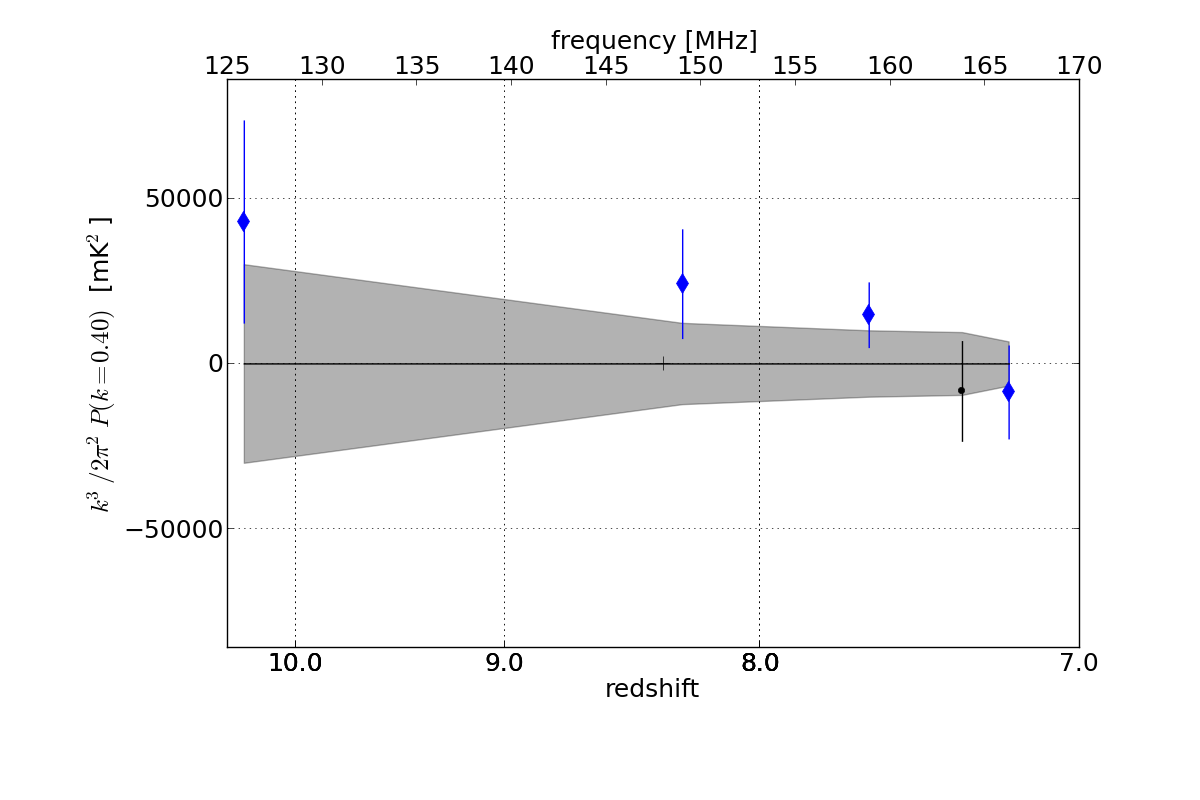}
\caption{Power spectrum amplitude vs redshift at a selection of k modes. Left to right from top, k=0.1,0.2,0.3,0.4 hMpc$^{-1}$.  Parsons 2014 PAPER limit marked with thin black, this work marked with thick blue diamonds.  The k=0.1 hMpc$^{-1}$ bin (top-left), which samples the delay spectrum at only 2x the maximum horizon delay  is foreground dominated with  a redshift dependence similar to the  $T_{rms}$ residual in Figure \ref{fig:noise}.  Most have amplitudes at or slightly above the noise level (grey region), particularly at high redshifts suggesting that most are systematic limited. Calculating the theoretical noise requires making several approximations (see Figure \ref{fig:pspecs} and Section \ref{sec:pspecs}).   \label{fig:slices}}

\end{figure*}

The residual level is also illustrated in the $k$ slices of Figure \ref{fig:slices}.  
 The shortest $k$ mode (upper left of Figure \ref{fig:slices}) is the nearest to the light travel horizon and therefore most likely to be contaminated by foreground leakage. This $k$ mode is well above the noise and has a redshift dependence much the same shape as the $T_{rms}$ curve in Figure \ref{fig:noise}, with a minimum near redshift 8.5 and a dramatic rise above redshift 10. These points are also the only ones where error bars significantly deviate from their predicted thermal values. As we see in Table \ref{tab:data}, going by predicted noise alone points closest to the horizon have error bars in excess of 20 $\sigma$ while the bootstrap error bars register a much larger uncertainty.  This hints at a more complicated origin for the excess. If these points were purely foregrounds, the error bars would be noise dominated, the fact that the excess is not much larger then its variation suggests a source of error not accounted in our analysis. Likely culprits include incomplete flagging of interference, instrumental glitches, or time of frequency dependent calibration variation which introduces differences between redundant baselines. All of which have are part of ongoing analysis efforts which are briefly discussed in the conclusions.   All of these effects have the effect of adding time and frequency dependence to the bright foregrounds limiting the effectiveness of the filter.   Thus, just like foregrounds, the strong excess drops rapidly towards higher $k$ modes. Above $k$ of 0.1 hMpc$^{-1}$  the power spectra are generally positively biased, though most close to 2 $\sigma$ or less using either theoretical and bootstrap error bars.

\begin{deluxetable*}{llccccccc}
\tablecolumns{9}
\tablecaption{PAPER Power spectrum values using 32 antenna and a full season of integration.}
\tablehead{
\colhead{k} & 
\colhead{redshift}  & 
\colhead{$\Delta^2$} &
\colhead{\parbox{6em}{theoretical error} }&
\colhead{\parbox{6em}{theoretical significance}} &
\colhead{\parbox{6em}{theoretical detection}} &
\colhead{\parbox{6em}{bootstrap error }} &
\colhead{\parbox{6em}{bootstrap significance}} &
 \colhead{\parbox{5em}{bootstrap detection}} \\
\colhead{[hMpc$^{-1}$]} & 
\colhead{}  & 
\colhead{[mK$^2$]} &
\colhead{[mK$^2$]} &
\colhead{ $\sigma$} & 
\colhead{}&
\colhead{\parbox{7em}{[mK$^2$]} }& 
\colhead{$\sigma$} & 
\colhead{}
}
\startdata
0.1 & 10.29 & 43600 & $\pm$   400 & 105.6 & Det & $\pm$  5700 & 7.6 & Det \tabularnewline
0.1 & 8.54 &  5700 & $\pm$   200 & 33.0 & Det & $\pm$  1200 & 4.9 & Det \tabularnewline
0.1 & 7.94 &  7700 & $\pm$   100 & 56.2 & Det & $\pm$  2000 & 3.8 & Det \tabularnewline
0.1 & 7.55 &  9500 & $\pm$   100 & 122.7 & Det & $\pm$  2800 & 3.4 & Det \tabularnewline
0.2 & 10.29 & 11400 & $\pm$  3300 & 3.5 & Det & $\pm$  5000 & 2.3 & Det \tabularnewline
0.2 & 8.54 &  5900 & $\pm$  1400 & 4.3 & Det & $\pm$  2200 & 2.7 & Det \tabularnewline
0.2 & 7.94 &  5100 & $\pm$  1100 & 4.8 & Det & $\pm$  1800 & 2.9 & Det \tabularnewline
0.2 & 7.55 &   900 & $\pm$   600 & 1.4 & Det & $\pm$  1500 & 0.6 & ULim \tabularnewline
0.3 & 10.29 & 34800 & $\pm$ 11000 & 3.2 & Det & $\pm$ 16400 & 2.1 & Det \tabularnewline
0.3 & 8.54 & 12900 & $\pm$  4600 & 2.8 & Det & $\pm$  5400 & 2.4 & Det \tabularnewline
0.3 & 7.94 &  5200 & $\pm$  3600 & 1.4 & Det & $\pm$  4800 & 1.1 & Det \tabularnewline
0.3 & 7.55 &  7600 & $\pm$  2000 & 3.8 & Det & $\pm$  4800 & 1.6 & Det \tabularnewline
0.4 & 10.29 & 42900 & $\pm$ 26000 & 1.7 & Det & $\pm$ 30700 & 1.4 & Det \tabularnewline
0.4 & 8.54 & 24100 & $\pm$ 10800 & 2.2 & Det & $\pm$ 16600 & 1.5 & Det \tabularnewline
0.4 & 7.94 & 14700 & $\pm$  8500 & 1.7 & Det & $\pm$  9900 & 1.5 & Det \tabularnewline
0.4 & 7.55 & -8600 & $\pm$  4800 & -1.8 & Det & $\pm$ 14200 & -0.6 & ULim \tabularnewline
\enddata
\tablenotetext{a}{All error bars are 1 $\sigma$.}
\tablenotetext{b}{Det indicates a measurement and error-bar inconsistent with zero, ULim indicates consistency with zero at 1 $\sigma$.}
\label{tab:data}
\end{deluxetable*}

\section{Conclusions}
\label{sec:conclusion}

The power of the highly redshifted 21 cm as a cosmic probe lies in its ability to probe a 3D volume by observing at different frequencies.  The present analysis extends the work of  \cite{Parsons:2014p10499}, which used a single-baseline delay spectrum analysis to place an upper limit on the 21 cm power at z = 7.68 (163 MHz).  In this work, we take advantage of the wide bandwidth of the PAPER instrument to place limits on the power spectrum at a range of redshifts $10.7>z>7.2$.  

Most of the power spectrum data points demonstrate a removal of foreground signals to approximately twice the thermal noise limit. The deepest point --at redshift 7.55 and $k=0.2$-- is a 1$\sigma$ upper limit of (48mK)$^2$. For comparison, the \citet{Lidz:2008p8251} fiducial model of the reionization power spectrum estimates an amplitude of (4.4mK)$^2$, for models with 50\% reionization at $z=$7.55.  Though most points are apparent detections of residual foregrounds they are still at a much deeper limit then previously found at these redshifts.

The fact that the best upper limit still comes from the z = 7.6 band does illustrate that this redshift corresponds to a somewhat special frequency for the PAPER instrument, where the combined contribution of system noise, RFI signals, and residual foregrounds (if any remain) are at a minimum.  The success of the filtering style foreground removal heavily leverages the surrounding spectral coverage, and is somewhat less effective at removing foregrounds across the entire band. 

Further work will look to expand on this result on several fronts.  Alternative algorithms to the wide-band CLEAN could potentially better remove the delay-space covariance introduced by RFI flags in frequency and are under investigation.  Another active point of investigation is aimed at alternative windows to the Blackman-Harris used in the wide-band CLEAN (or wholly distinct algorithms) that could better remove foregrounds from the edges of the band while still limiting foreground bleed into the EoR window.  Weighting of the final power spectrum average is also a subject for further improvement.   In this analysis we have used uniform weights; this is also the default assumption of most sensitivity calculations, including the one used here.  Uniform weighting is optimal if the final measurements are limited by something other than noise, while inverse variance is appropriate for purely noise limited measurements.  Given how close the error bars are to noise dominated, future analysis will investigate using different estimates of the variance for optimal weighting.

Future work will also include observations from subsequent seasons with more antennas.  The observations reported here demonstrate a nearly noise limited integration over a full season but used only a quarter of the final design antenna count. These  32 antenna observations from 2011 were followed by a full season of 64 antennas in 2012 and 128 in 2013.  A second season of observing with the 128 antenna array is now under way. In going from  32 to 128 antenna the mK$^2$ sensitivity increases by the expected factor of 4 and by an additional factor of $\sim$2 after accounting for the substantial uv-plane redundancy \citep{Parsons:2012p9028}. 

The analysis of the system temperature presented in section 4.1 is one of the most thorough such investigations applied to the PAPER instrument.  It confirms the effectiveness of the wideband CLEAN in removing foregrounds over a wide range of the band, while illustrating weaknesses in this approach in pushing towards the edges of PAPER's frequency range.  Furthermore, after the application of the fringe rate-based time averaging, we see hints that baseline-to-baseline variations between redundantly spaced baselines are becoming the dominant systematic in the analysis.  Techniques like the covariance removal introduced in P14 and also used in this work can differentiate some of this variation from the sky signal; however some combination of improved calibration techniques \citep{Zheng:2014p10467}, model building \citep{Sullivan:2012p9457}, and a better understanding of error covariance  \citep{Liu:2011p8763} may also prove valuable in reducing this systematic. Ongoing analysis of more sensitive data sets suggests that the excess seen here can be minimized further with some combination of: fringe rate filters that are better matched to the sky and time dependent calibration (Ali et al, in prep).

\section{Acknowledgements}
The PAPER project is supported through the NSF-AST program (award \#1129258). Computing resources were provided by a grant from Mt. Cuba Astronomical Foundation.  J.C.P. is supported by an NSF Astronomy and Astrophysics Fellowship under award AST \#1302774.   We thank Paul Sutter and Jonathan Pritchard for helpful discussions.

\end{document}